\documentstyle[12pt,twoside,epsfig]{article}
\begin{document}
\title{\Large \bf Recent Results from BES}

\author{Xiaoyan Shen\\
        Representing the BES Collaboration\\
	Institute of High Energy Physics, Academy of Sciences,\\
	Beijing 100039, P.R. China}
\date{}
\maketitle     
\begin{abstract}
We report partial wave analysis results for 
$J/\psi \to \gamma K^+ K^-$ based on $7.8 \times 10^6$ BESI 
$J/\psi$ events, and find $0^{++}$ to be dominant in the $f_J(1710)$
mass region. Some very preliminary results are presented from the
$2.2 \times 10^7$ $J/\psi$ events newly collected at the upgraded BES(BESII). 
Using the
world largest $\psi(2S)$ data sample, BES measures branching 
ratios of $\psi(2S)$ radiative and hadronic decays.
The preliminary R values measured by BESII in the 2-5 GeV energy region
are also presented.
\end{abstract}

\section{Introduction}
BES is a large general purpose solenoidal detector at the Beijing Electron
Positron Collider (BEPC). The details of BESI are described in 
ref. \cite{BES}.
The upgrades of BESI to BESII include the replacement of the central drift
chamber with a vertex chamber composed of 12 tracking layers, the
installation of a new barrel time-of-flight counter (BTOF) with a 
time resolution of 180ps and the installation of a new main 
drift chamber (MDC), which has 10 tracking layers and provides
a $dE/dx$ resolution of $\sigma_{dE/dx} = 8.4\%$ for particle 
identification and $\sigma_p/p = 1.8\% \sqrt{1+p^2}$ ($p$ in GeV) momentum
resolution for charged tracks. The barrel shower counter
(BSC), which covers $80\%$ of $4\pi$ solid angle, has an energy resolution 
of $\sigma_E/E = 23\%/\sqrt{E}$ ($E$ in GeV) and a spatial resolution 
of 7.9 mrad in $\phi$ and 2.3 cm in z, is located outside the TOF. 
Outermost is a $\mu$ identification system, which consists of three
double layers of proportional tubes interspersed in the iron flux return
of the magnet.

\section{\boldmath $J/\psi$ physics}
\subsection{\boldmath $J/\psi$ data sample}
\bigskip
Based on a $7.8 \times 10^6$ $J/\psi$ event sample, collected at BESI, 
many studies
on $J/\psi$ decays have been performed and some are published. At the
end of 1999, we started a new $J/\psi$ run with the upgraded BESII.
Up to now we have accumulated $2.2 \times 10^7$ $J/\psi$ events, which is
already the largest $J/\psi$ sample in the world. By the end of 2001,
we hope to reach our goal of collecting $5 \times 10^7$ $J/\psi$ events.
Fig. 1 shows us the $J/\psi$ event samples in the world.   

\begin{figure}[htb]
\centerline{\hbox{\psfig{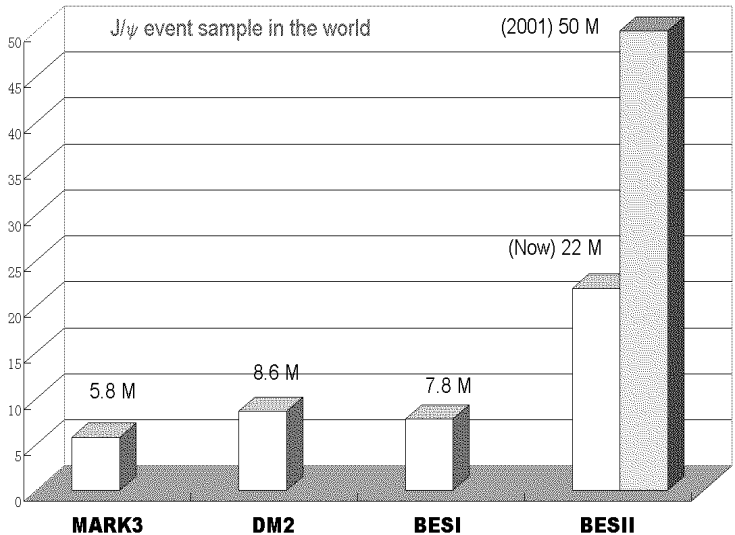}}}
\caption{$J/\psi$ event samples in the world}
\label{fig:jpsi_sample}
\end{figure}

\subsection{\boldmath Recent results from BESI $J/\psi$ data}
\bigskip
One of the distinctive features of QCD as a non-Abelian gauge theory is the
self-interaction of gluons. The indirect evidence for gluon-gluon
interactions has been obtained at high energies. However, glueballs,
the bound states of gluons, predicted by QCD, have
not been confirmed yet. Therefore, the observation
of glueballs is, to some extent, a direct test of QCD. The  
$f_J(1710)$, first observed by the Crystal Ball Collaboration
in $J/\psi \to \gamma \eta \eta$ \cite{cbl}, has been considered as the 
lightest $0^{++}$ glueball candidate because of its large production 
rate in gluon 
rich processes, such as $J/\psi$ radiative decays, $p p$ central production
{\it{etc}}, and because of the lattice QCD calculation of the lightest $0^{++}$ glueball
mass \cite{gluem}. However, the spin-parity of $f_J(1710)$ is not 
determined after many
years' efforts. Based on BESI $7.8 \times 10^6$ $J/\psi$ data, a partial
wave analysis is performed to the $f_J(1710)$ mass region in 
$J/\psi \to \gamma K^+ K^-$ channel. In each of the subplots in Fig. 2,
$K^+ K^-$ invariant mass is depicted as points with error bars, and the 
resulting components from fits to the data are shown as the solid 
histograms.

\begin{figure}
\centerline{\hbox{\psfig{file=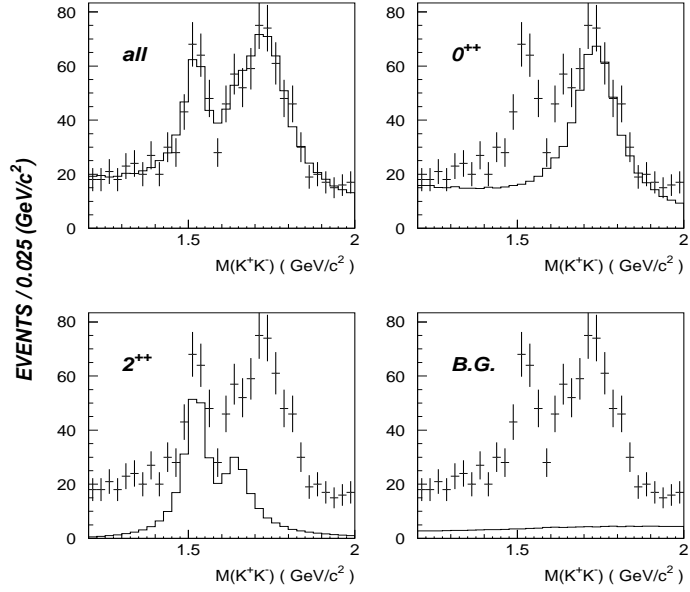,height=90mm,width=100mm}}}
\caption{Projection of the different components on the $K^+ K^-$ mass}
\label{fig2:fj_proj}
\end{figure}

\subsection{\boldmath BESII $J/\psi$ data}
\bigskip
At present, the newly accumulated $2.2 \times 10^7$ $J/\psi$ data has been 
calibrated and reconstructed. The inclusive $\phi$, $\Lambda$, $K^*$ and 
$K_s^0$ signals and their fitted masses are shown in Fig. 3.   

\begin{figure}
\centerline{\hbox{\psfig{file=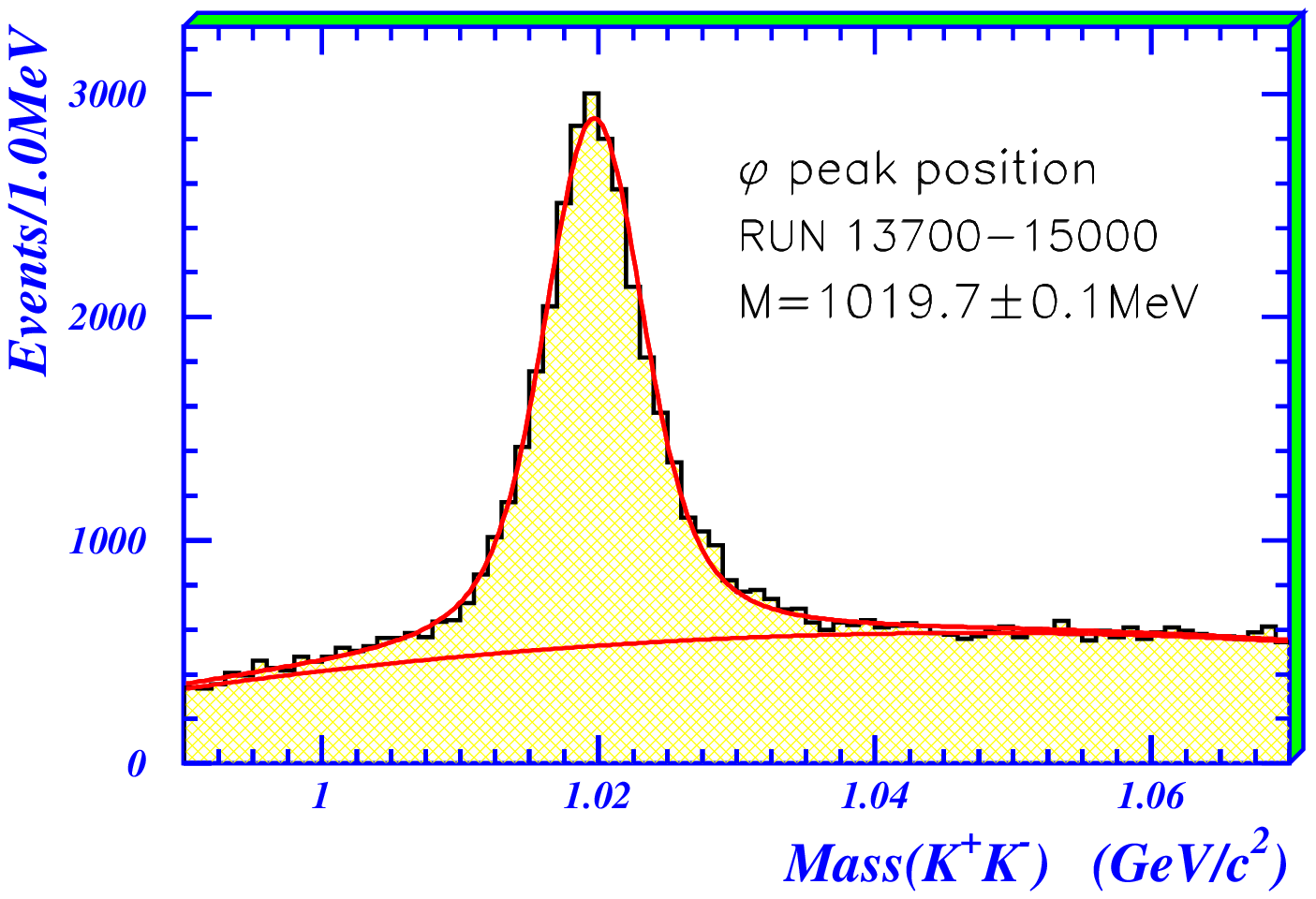,height=50mm,width=50mm}
\psfig{file=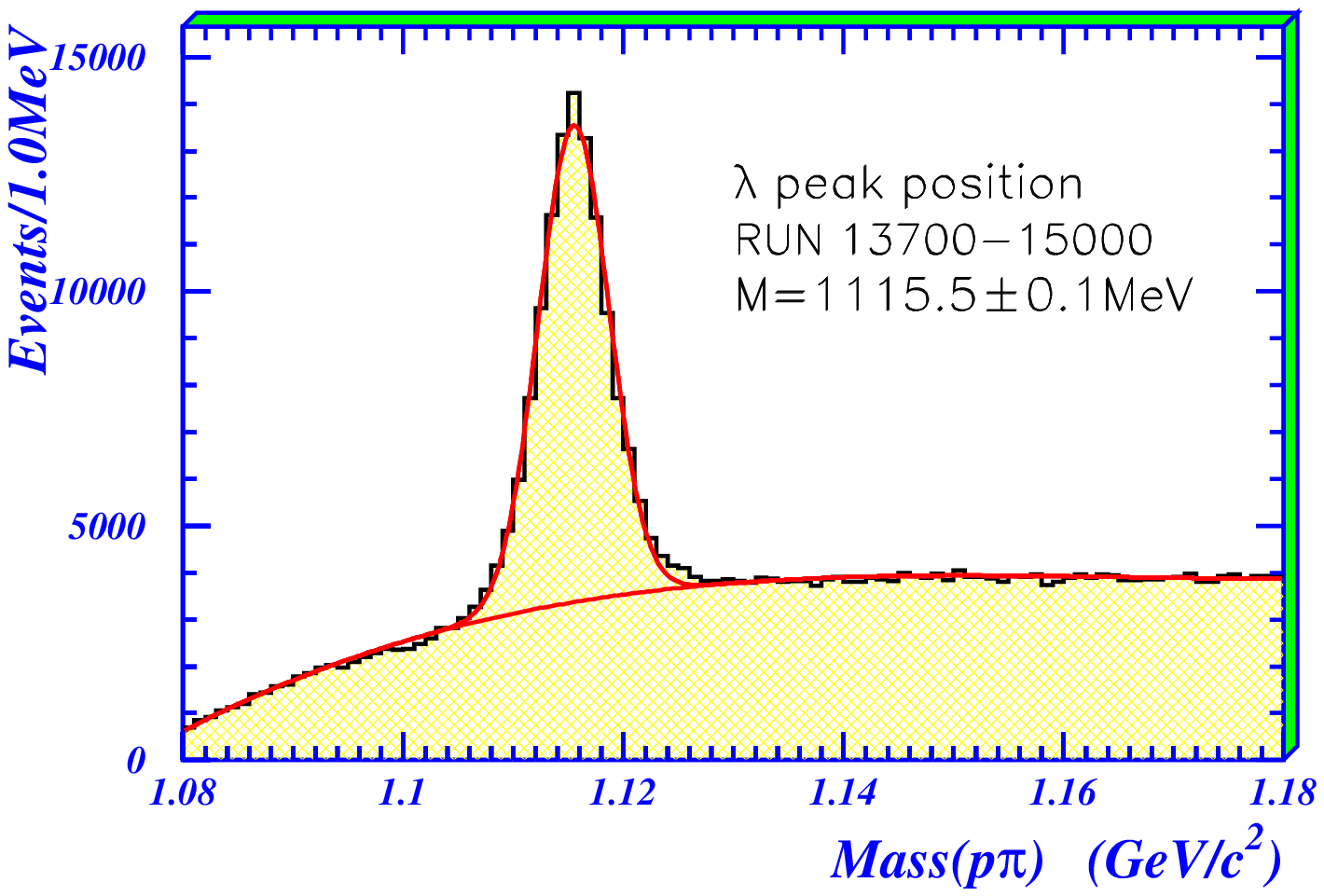,height=50mm,width=50mm}}}
\centerline{\hbox{\psfig{file=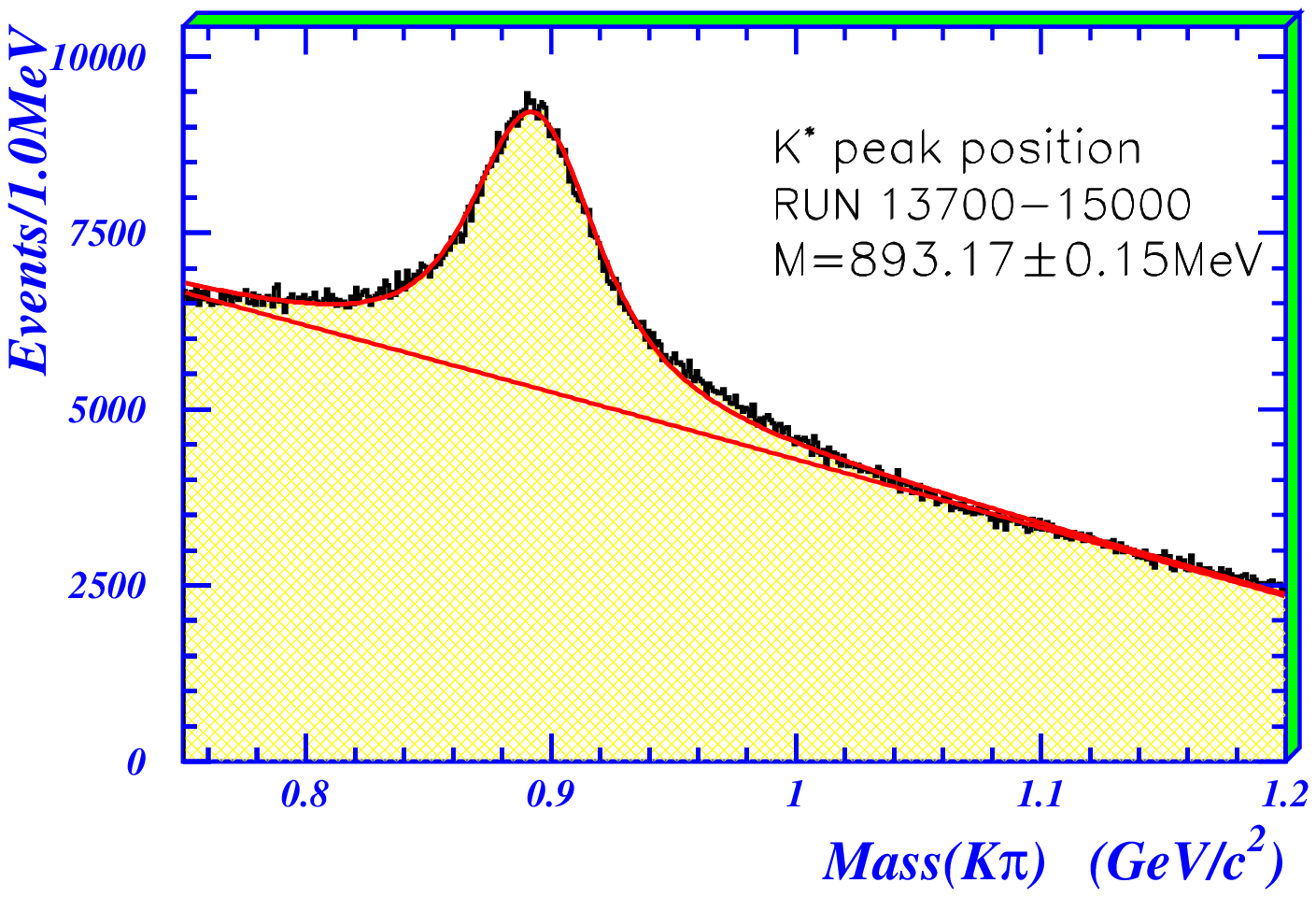,height=50mm,width=50mm}
\psfig{file=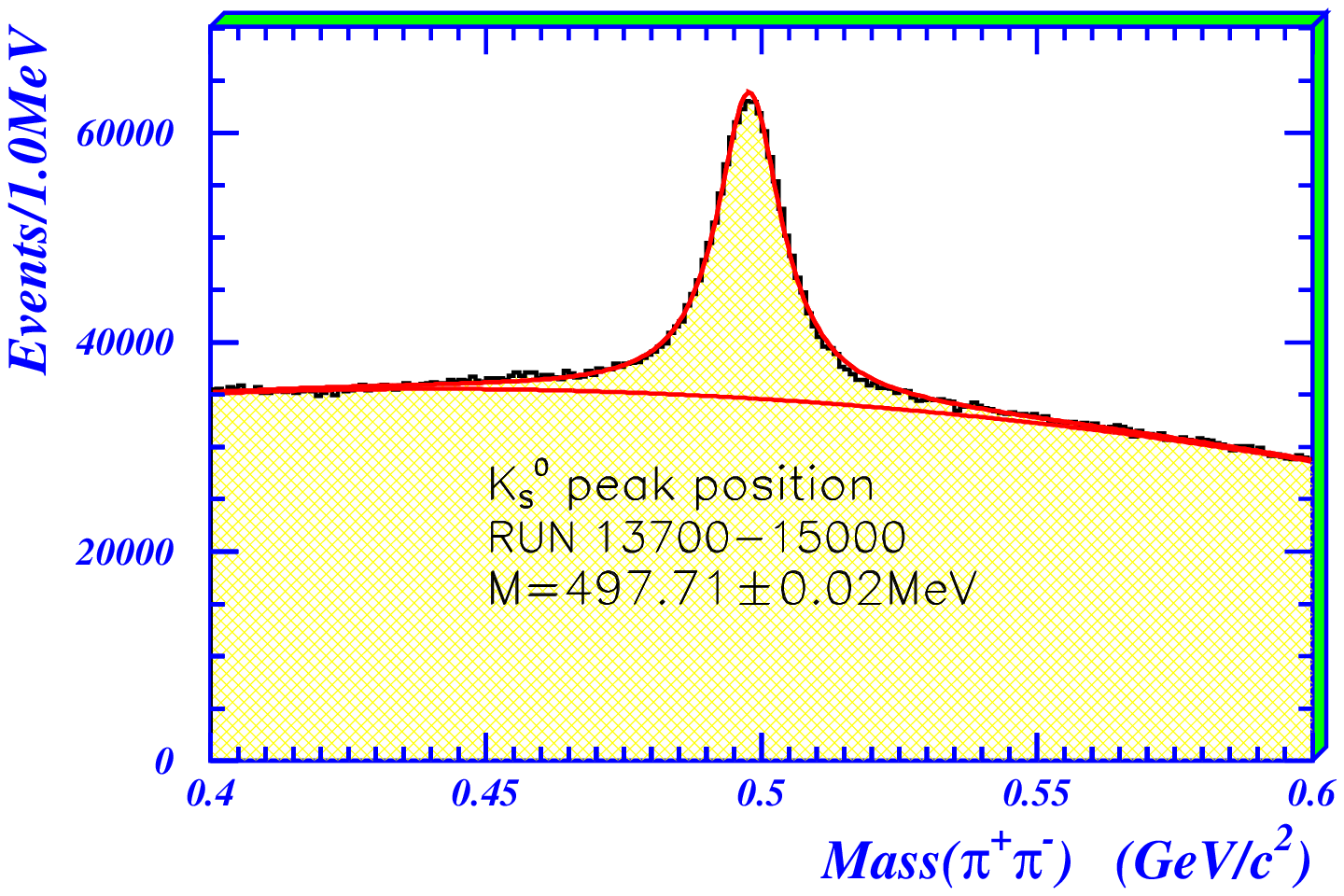,height=50mm,width=50mm}}}
\caption{Inclusive $\phi$, $\Lambda$, $K^*$ and $K_s^0$ signals}
\end{figure}      

Fig. 4 is the invariant mass spectra and Dalitz plot for
$J/\psi \to \omega \pi^+ \pi^-$. Clear $\pi^0$ 
and $\omega$ signals are seen in the 2$\gamma$ and $\pi^+ \pi^- \pi^0$ 
invariant mass spectra, respectively. In the $\pi^+ \pi^-$ mass distribution, 
which recoils against the $\omega$, an $f_2(1270)$ 
and a big bump around 500 MeV 
are observed. The invariant masses of $M_{\pi^+ \pi^-}$ and
$M_{K \pi}$ are plotted in Fig. 5 for $J/\psi \to K^{*\pm}K^{\mp}$, 
$K^{*\pm} \to K_s^0 \pi^{\pm}$ decay. Both $K_s^0$ and $K^*$ are nicely peaked.

\begin{figure}
\centerline{\hbox{\psfig{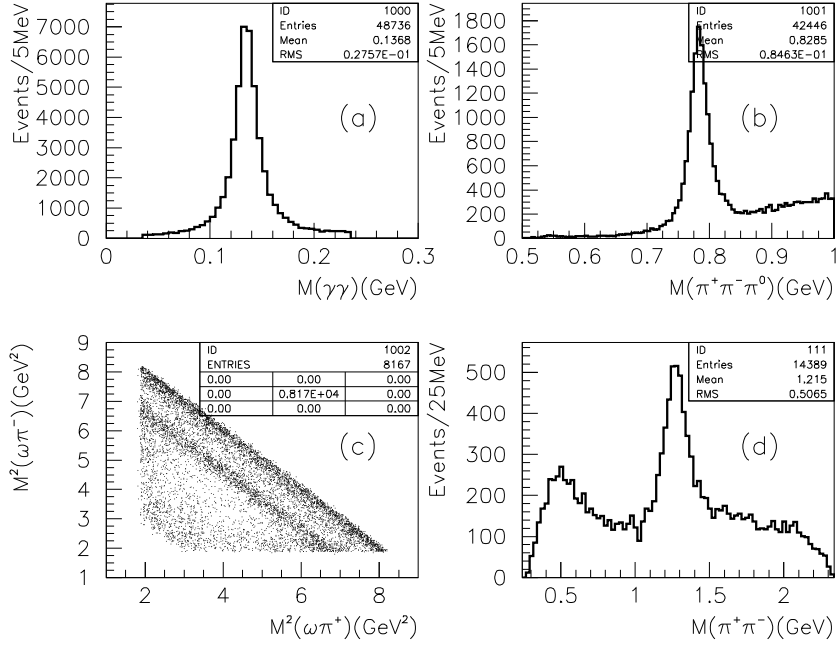}}}
\caption{$J/\psi \to \omega \pi^+ \pi^-$, $\omega \to
\pi^+ \pi^- \pi^0$ decay. (a) Invariant mass of $\gamma \gamma$. (b)
Invariant
mass of $\pi^+ \pi^- \pi^0$. (c) Daliz plot. (d) Invariant mass of $\pi^+
\pi^-$}
\label{fig4:wpipi}
\end{figure}

\begin{figure}
\centerline{\hbox{\psfig{file=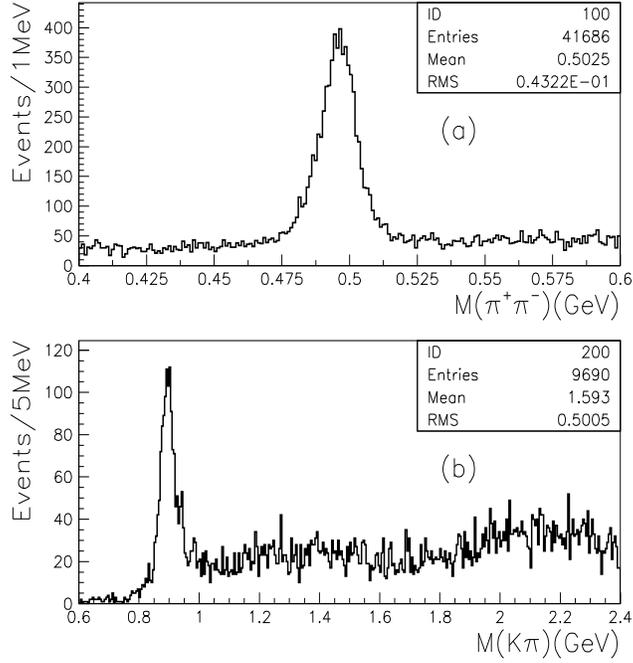,height=100mm,width=90mm}}}
\caption{$J/\psi \to K^{*\pm}K^{\mp}$, $K^{*\pm} \to K_s^0
\pi^{\pm}$, $K_s^0 \to \pi^+ \pi^-$ decay. (a) Invariant mass of
$\pi^+ \pi^-$. (b) Invariant mass of $K \pi$}
\label{fig5:Kstark}
\end{figure}         

\subsection{\boldmath Preliminary results from BESII $J/\psi$ data}
\bigskip
Based on the $2.2 \times 10^7$ $J/\psi$ events, some
very preliminary results are obtained.  

\subsubsection{\boldmath PWA analyses on $J/\psi \to \phi \pi^+ \pi^-$ and $\phi K^+
K^-$}

\bigskip
Partial Wave Analyses (PWA) of $J/\psi \to \phi \pi^+ \pi^-$ 
and $\phi K^+ K^-$ are performed.
Fig. 6 represents the contribution of every component from the fit to 
$J/\psi \to \phi \pi^+ \pi^-$. Three $0^{++}$ resonances, located at 
980 MeV, 1370 MeV and 1770 MeV, and one $2^{++}$ resonance at 1270 MeV 
are observed 
in the $\pi^+ \pi^-$ invariant mass recoiling against the $\phi$. 
In $J/\psi \to \phi K^+ K^-$, $f_2'(1525)$ and $f_0(1710)$ components 
are found to be needed for a good fit. The projection of different 
components in $K^+ K^-$ mass is shown in Fig. 7.

\begin{figure}
\centerline{\hbox{\psfig{file=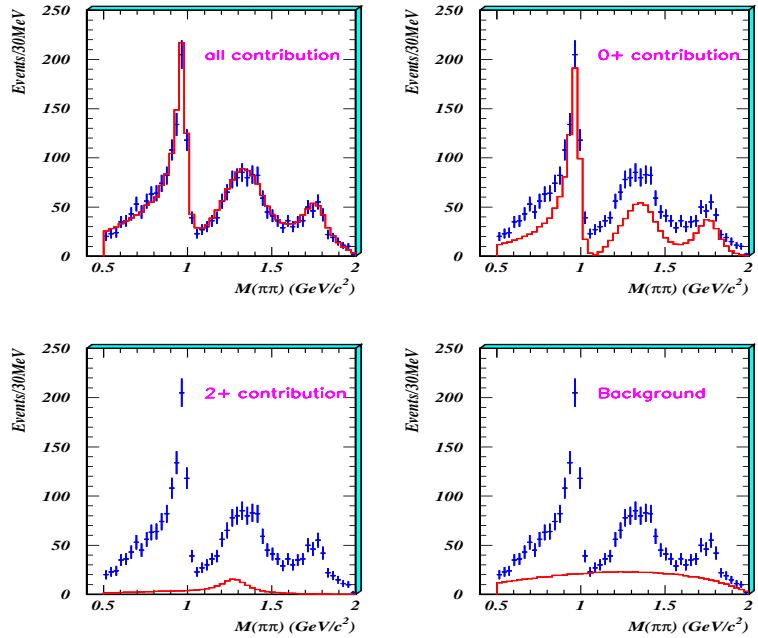,height=90mm,width=110mm}}}
\caption{Projection of every component on $\pi^+ \pi^-$ mass in $J/\psi \to
\phi \pi^+ \pi^-$. Points with error bars are data, and the solid histograms
represent the fit curves (Very preliminary).}
\label{fig7:phipipi}
\end{figure}

\begin{figure}
\centerline{\hbox{\psfig{file=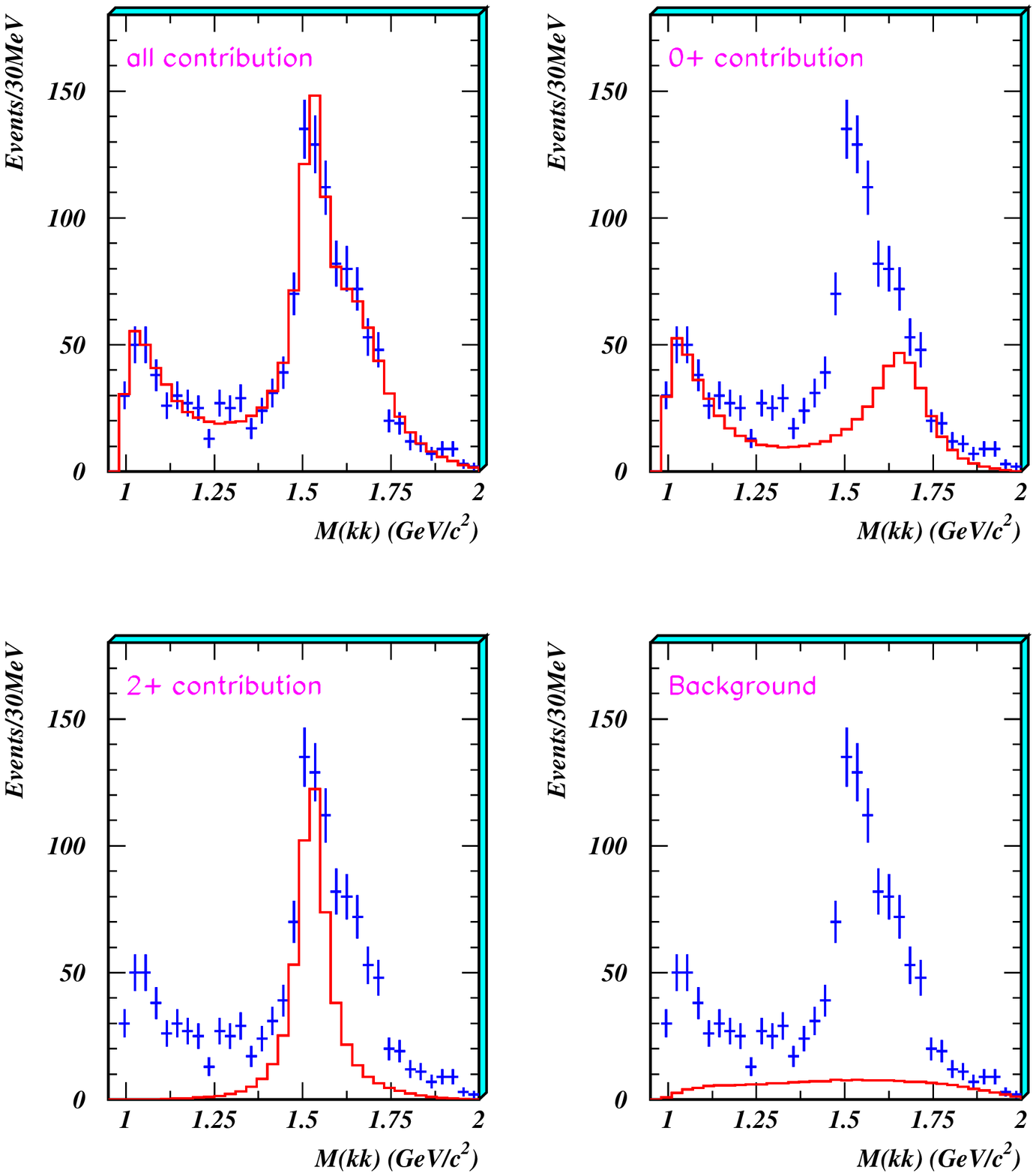,height=90mm,width=110mm}}}
\caption{Projection of every component on $K^+ K^-$ mass in $J/\psi \to \phi
K^+ K^-$. Points with error bars are data, and the solid histograms
represent the fit curves  (Very preliminary).}
\label{fig8:phikk}
\end{figure}          

\subsubsection{\boldmath Inclusive $\gamma$ spectrum}

\bigskip
$J/\psi$ radiative decay $J/\psi \to \gamma X$ is a rich source of 
glueballs. In addition to studying the radiative decays 
exclusively, the inclusive $\gamma$ spectrum is another place to 
search for glueballs. Due to the relatively poor energy 
resolution for BSC, we use $\gamma$ conversion to 
$e^+ e^-$ pairs inside our detector and then measure the momenta of $e^+$
and $e^-$ in the MDC, which has a momentum resolution of $1.8\% \sqrt{1+p^2}$
($p$ in GeV), to avoid using BSC energy information and thus improve the 
energy resolution. 
Fig. 8 
plots the energy resolution of $E_{e^+e^-}$ via converted $\gamma$ energy. 
The inclusive $\gamma$ spectrum is plotted in Fig. 9.
One can see from Fig. 9 that there is a bump at $E_{\gamma}=0.75$ GeV, 
which corresponds to the position of $\xi(2230)$. The energy resolution is
about 12 MeV at that point.

\begin{figure}
\centerline{\hbox{\psfig{file=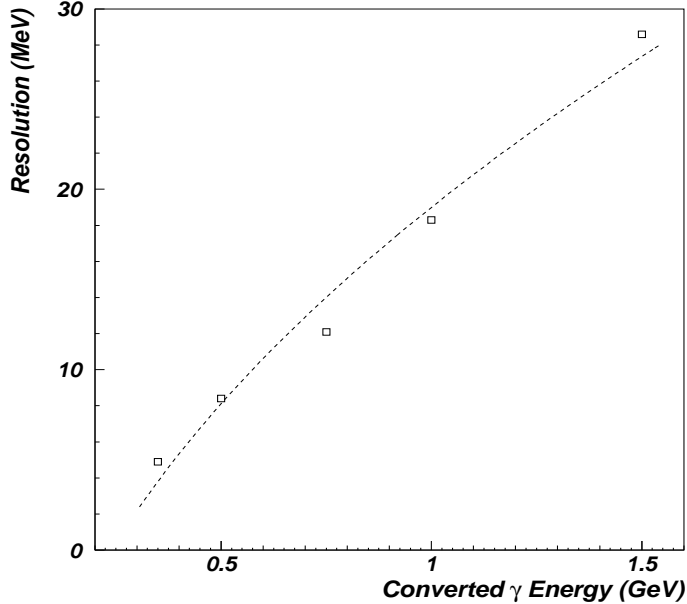,height=80mm,width=90mm}}}
\caption{Energy resolution of $E_{e^+e^-}$ via converted $\gamma$ spectrum
(Very preliminary)}
\label{fig8:resolution}
\end{figure}

\begin{figure}[htb]
\centerline{\hbox{\psfig{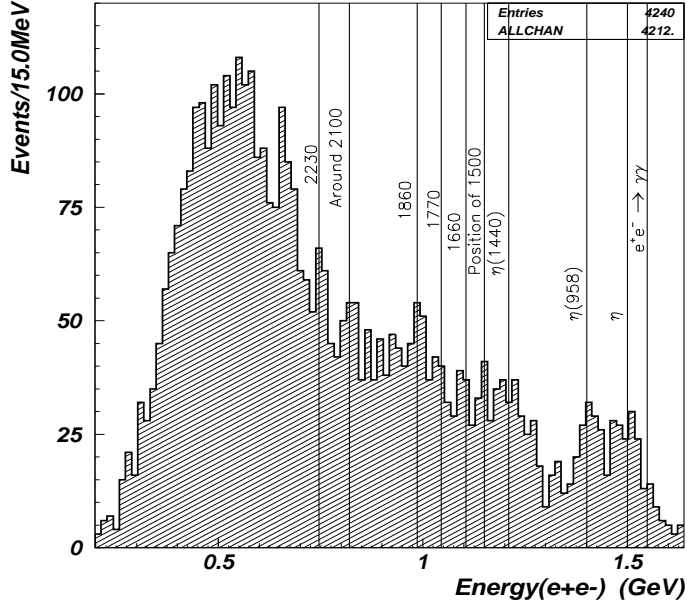}}}
\caption{Inclusive $\gamma$ ($E_{e^+e^-}$) spectrum (Very preliminary).}
\label{fig9:inclusive_g}
\end{figure}       

\section{\boldmath $\psi(2S)$ physics}

\subsection{\boldmath $\psi(2S)$ hadronic decays}
\bigskip
According to perturbative QCD, the dominant process for hadronic decays 
of both 
$J/\psi$ and $\psi(2S)$ is the annihilation of 
$c$ and $\bar c$ quark into three gluons. 
In this case, the partial width for the decay is 
proportional to the wave function at the origin in the non-relativistic 
quark model for $c\bar{c}$. Therefore one can expect~\cite{sjb} 
that, for any hadronic final state $h$,
$$Q_{h}\equiv\frac{B(\psi(2S)\rightarrow h)}{B(J/\psi\rightarrow h)} 
=\frac{B(\psi(2S)\rightarrow e^{+}e^{-})}{B(J/\psi\rightarrow
e^{+}e^{-})}=0.148\pm0.022 $$.\\ 
\noindent This is the so called ``$15\%$'' rule.\\

In order to test ``$15\%$'' rule, systematic studies on $\psi(2S)$ hadronic
decays have been carried out based on $3.96 \times 10^6$ $\psi(2S)$ events,
which was collected by BESI and is the largest $\psi(2S)$ data sample in 
the world. Table 1 summarizes some of the branching ratios of
$\psi(2S)$ hadronic
decays, that we have measured so far, as well as the ratios $Q_h$ of 
branching fractions of $\psi(2S)$ to $J/\psi$. Of those branching fractions
of $\psi(2S)$ listed in Table 1, many are measured for the first time and 
many improve over PDG values.\\

\begin{table}
\begin{center}
\caption{Branching fractions of $\psi(2S)$ decays and ''$15\%$'' rule
test (limits are at C.L.$90\%$)}
\begin{tabular}{lll}
\hline
 Process        &  $B(\times 10^{-5})$ &$Q_h$ ($\%$)\\ \hline      

$\omega K^+K^-$         &  $12.5\pm 5.6$  &  $16.9\pm 9.4$ \\\hline

$\omega p\overline{p}$  &  $6.4\pm 2.6$   &  $5.0\pm 2.2$  \\\hline

$\phi\pi^+\pi^-$        &  $16.8\pm 3.2$  &  $21.0\pm 5.1$ \\\hline

$\phi K^+K^-$           &  $5.8\pm 2.2$   &  $7.0\pm 2.9$  \\\hline

$\phi p\overline{p}$    & $0.82\pm 0.52$ & $18.1\pm 12.8$ \\\hline

$\phi f_0$              &  $6.3\pm 1.8$   & $19.6\pm 7.8$  \\\hline

$ K^*K^-\pi^+ +c.c.$    &  $60.4\pm 9.0$  &  ?             \\\hline

$ K^*\overline{K}^*+c.c.$ & $3.92\pm 1.03$ & $13.6\pm 4.9$\\\hline

$K^*\overline{K_2}^*+c.c.$ & $7.98\pm 5.28$ &$1.20\pm 0.93$ \\\hline

$\pi^0\pi^+\pi^-p\overline{p}$& $34.9\pm 6.4$ & $15.2\pm 6.6$ \\\hline  

$\eta\pi^+\pi^-p\overline{p}$ & $24.7\pm 9.6$   &  ? \\\hline

$\eta p\overline{p}$      & $<18. $             & $ <8.6$   \\\hline       

$p\overline{p}$    &$21.6 \pm 3.9$ & $10.1 \pm 1.9$ \\\hline

$\Lambda\overline{\Lambda}$ &$18.1 \pm 3.4$ & $13.4 \pm 2.9$ \\\hline

$\Sigma^0\overline{\Sigma}^0$  &$ 12 \pm 6$ & $9.4 \pm 4.6$ \\\hline

$\Delta^{++}\overline{\Delta}^{--}$ & $12.8 \pm 3.5$ & $11.6 \pm 4.5$ \\\hline

$\Xi^-\overline{\Xi}^+$		& $9.4 \pm 3.1$  &$ 10.4 \pm 4.1$ \\\hline

$\Sigma^{*-}\overline{\Sigma}^{*+}$   & $11 \pm 4$ &$ 11 \pm 4$ \\\hline

$\Xi^{*0}\overline{\Xi}^{*0}$       &$<8.1 $   &     \\\hline

$\Omega^-\overline{\Omega}^+$       &$<7.3$    &     \\\hline

\end{tabular}
\end{center}
\end{table}

\subsection{\boldmath $\psi(2S)$ radiative decays}
\bigskip 
Table 2 lists the results on $\psi(2S)$ radiative decays. The branching 
fractions for $\psi(2S) \to \gamma f_2(1270)$ and $\psi(2S) \to \gamma
f_J(1710) \to \gamma K \overline K$ agree with ''$15\%$'' rule, and the 
ratio of $\chi_{c0} \to \eta\eta$ to $\chi_{c0}\to \pi^0\pi^0$ ($0.73
\pm 0.39$), is consistent with the theoretical prediction of 0.95 
based on flavor SU(3) symmetry.

\begin{table}
\begin{center}
\caption{Branching ratios of $\psi(2S)$ radiative decays}  
\begin{tabular}{ll}
\hline
 Process        &  $B(\times 10^{-4})$        \\ \hline
$\gamma f_2(1270)$  & $2.27\pm0.43$ \\ \hline
$\gamma f_J(1710)\to \gamma\pi\pi$ & $0.336\pm0.165$   \\ \hline
$\gamma f_J(1710)\to \gamma K^+K^-$ & $0.55\pm0.21$   \\ \hline
$\gamma f_J(1710)\to \gamma K^0_S K^0_S$ & $0.21\pm0.15$   \\ \hline
$\gamma \chi_{c0}\to \gamma \pi^0\pi^0$  & $26.8 \pm 6.5$ \\ \hline
$\gamma \chi_{c2}\to \gamma \pi^0\pi^0$  & $8.8\pm 5.6$ \\ \hline
$\gamma \chi_{c0}\to \gamma\eta\eta$  & $19.4\pm 10.0$ \\ \hline
$\gamma \chi_{c2}\to \gamma \eta\eta$  & $ < 12.2$ $(90\%$ C.L.) \\ \hline
\end{tabular}
\end{center}
\end{table}

\subsection{\boldmath $\psi(2S)$ resonance parameters}
\bigskip
To determine $\psi(2S)$ decay width, BES scanned 24 energy points,
in 3.67 to 3.71 GeV region, with an integrated
luminosity around $790 nb^{-1}$. The cross sections for $\psi(2S)
\to$ hadrons, $\pi^+ \pi^- J/\psi$ and $\mu^+ \mu^-$ are shown in Fig. 10.
The solid curves are the fit curves to the data. The final results
and systematic errors are still under study.  

\begin{figure}
\centerline{\hbox{\psfig{file=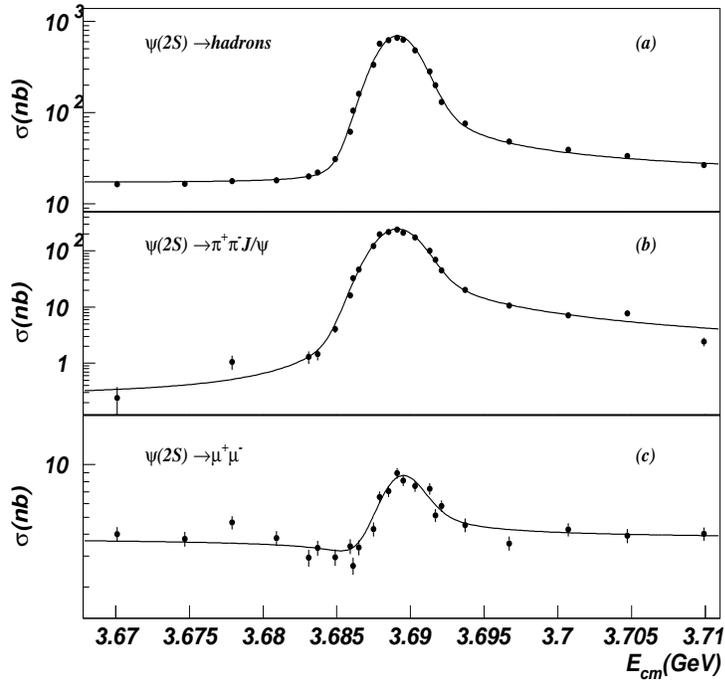,height=90mm,width=95mm}}}
\caption{The cross sections for $\psi(2S) \to hadrons$, $\pi^+ \pi^- J/\psi$
and $\mu^+ \mu^-$. Solid curves are fit curves (preliminary).}
\label{fig11:psip}
\end{figure}    

\section{R measurement at BES}

The QED running coupling constant $\alpha(s)$ and the anomalous magnetic moment
of muon $a_{\mu}$ are two fundamental quantities for the precision test of 
the Standard Model \cite{blondel,zhao}. The precisely measurement of R, which 
is defined as: 
$$ R = \frac{\sigma(e^+e^- \to hadrons)}{\sigma(e^+e^- \to \mu^+ \mu^-)},$$
is essential for interpreting $g-2$ experiment carried out at BNL and 
precisely evaluating $\alpha_{QED}(M_z^2)$.

Two scans were performed with BESII to measure R in the
energy region of 2-5 GeV in 1998 and 1999. The first run scanned 6
energy points covering the energy from 2.6 to 5 GeV in the continuum
and the results have been published \cite{besr}. The second run scanned 85 
points in the energy region of 2-5 GeV. The average uncertainty on R
is $7\%-10\%$, which is a factor of 2-3 improvement compared to 
the previous measurements. Fig. 11 shows the R values 
at each energy point. A careful study of the systematic
errors is still underway and all the systematic errors are
conservatively assigned to be $10\%$ temporarily. 

\begin{figure}
\centerline{\hbox{\psfig{file=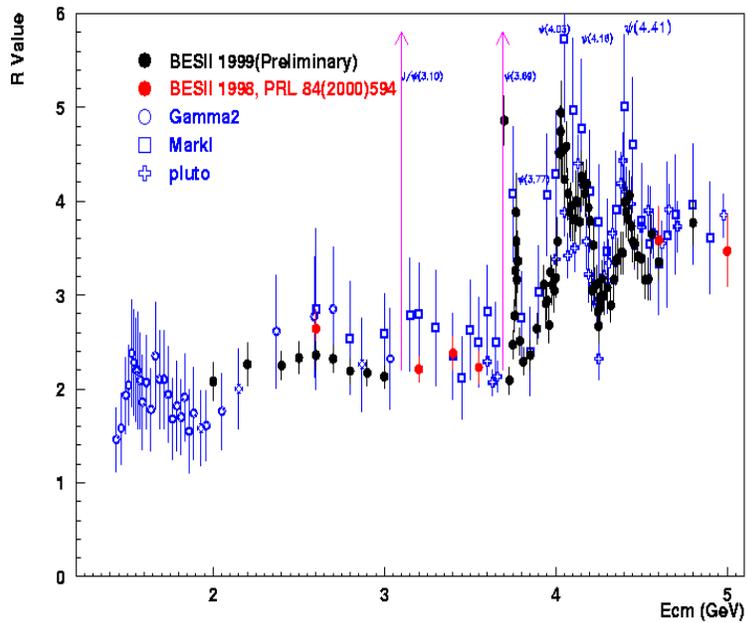,height=110mm,width=100mm,angle=-90.}}}
\caption{Plot of R values vs $E_{cm}$. All the systematic errors are
set to be $10\%$ (preliminary).}
\label{fig12:R_scan}
\end{figure}

\section{Summary}

Based on $7.8 \times 10^6$ BESI $J/\psi$ events, a partial wave analysis is 
applied to $J/\psi \to \gamma K^+ K^-$ decay, and $0^{++}$ is found 
to be dominant in the $f_J(1710)$ mass region. 

Since the end of 1999, $2.2 \times 10^7 J/\psi$ events have been
collected with BESII. Some very preliminary results are obtained. By the 
end of 2001, BESII will accumulate $5 \times 10^7 J/\psi$ events.

With $3.96 \times 10^6 \psi(2S)$ events, BES measured $\psi(2S)$ decay
branching fractions, many of them are measured for the first time 
and many improve on PDG values.

BES has scanned 6(1998)+85(1999) points in the energy region 
of 2-5 GeV since 1998. BES R measurement reduces the uncertainties on R 
from $15\%-20\%$ to $7\%-10\%$.  
 
\vspace{.4cm}
\noindent 
{\bf Acknowledgments}
\vspace{.1cm}

We acknowledge the staff of the BEPC accelerator and IHEP computing 
center for their efforts. The work was supported in part by the 
National Natural Science Foundation of China under Contracts No.19991480, 
No.19825116 and No.19605007, and by the Department of
Energy of US under Contracts No.DE-FG03-93ER40788 (Colorado State 
University), No.DE-AC03-76SF00515 (SLAC), No.DE-FG03-94ER40833 
(University of Hawaii) and No.DE-FG03-95ER40925 (University of
Texas at Dallas).


\begin{thebibliography}{100} 
\bibitem{BES} 
BES Collaboration, J. Z. Bai {\it et al.}, Nucl. Instrum.
Methods in Phys Res. Sect. A344 (1994) 319.
\bibitem{cbl} 
C. Edwards {\it et al.}, Phys. Rev. Lett., 56 (1986) 107.
\bibitem{gluem} 
C. Morningstar and M. Peardon, Phys. Rev. D60 (1999) 034509.
\bibitem{sjb} 
S. J. Brodsky and M. Karliner, Phys. Rev. Lett. {\bf 78} (1997) 468.
\bibitem{blondel} 
A. Blondel, Proc. of the 28th Int. Conf. on High
Energy Physics, Warsaw, Poland, 1996.
\bibitem{zhao}
Z. G. Zhao, Proc. of LP99, SLAC, USA, 1999.  
\bibitem{besr} 
BES Collaboration, J. Z. Bai {\it et al.}, Phys. Rev. Lett., 84 (2000) 594.
\end{thebibliography}
\end{document}